# Experimental determination of the valence band offsets of ZnGeN$_2$ and (ZnGe)$_{0.94}$Ga$_{0.12}$N$_2$ with GaN

Md Rezaul Karim[1], Brenton A. Noesges[2], Benthara Hewage Dinushi Jayatunga[3], Menglin Zhu[4], Jinwoo Hwang[4], Walter R. L. Lambrecht[3], Leonard J. Brillson[1,2], Kathleen Kash[3], Hongping Zhao[1,4,‡]

[1]Department of Electrical and Computer Engineering, The Ohio State University, Columbus, OH 43210, USA
[2]Department of Physics, The Ohio State University, Columbus, OH 43210, USA
[3]Department of Physics, Case Western Reserve University, Cleveland, OH 44106, USA
[4]Department of Materials Science and Engineering, The Ohio State University, Columbus, OH 43210, USA
‡ Email: zhao.2592@osu.edu

## Abstract

A predicted type-II staggered band alignment with an approximately 1.4 eV valence band offset at the ZnGeN$_2$/GaN heterointerface has inspired novel band-engineered III-N/ZnGeN$_2$ heterostructure-based device designs for applications in high performance optoelectronics. We report on the determination of the valence band offset between metalorganic chemical vapor deposition grown (ZnGe)$_{1-x}$Ga$_{2x}$N$_2$, for x = 0 and 0.06, and GaN using X-ray photoemission spectroscopy. The valence band of ZnGeN$_2$ was found to lie 1.45-1.65 eV above that of GaN. This result agrees well with the value predicted by first-principles density functional theory calculations using the local density approximation for the potential profile and quasiparticle self-consistent GW calculations of the band edge states relative to the potential. For (ZnGe)$_{0.94}$Ga$_{0.12}$N$_2$ the value was determined to be 1.29 eV, ~10-20% lower than that of ZnGeN$_2$. The experimental determination of the large band offset between ZnGeN$_2$ and GaN provides promising alternative solutions to address challenges faced with pure III-nitride-based structures and devices.

# 1. Introduction

ZnGeN$_2$ is the II-IV-N$_2$ analogue of wurtzite GaN in which half of the cation sites are occupied by Zn and the other half by Ge. Thermodynamically, the most stable phase of ZnGeN$_2$ has a wurtzite-derived orthorhombic Pna2$_1$ (also called Pbn2$_1$) space group. The band gap of this phase has been predicted to be within about 0.1 eV of that of GaN [1, 2]. Measurements of the band gap of ZnGeN$_2$ by photoluminescence and absorption spectroscopy are consistent with this prediction [3-5]. In addition, ZnGeN$_2$ is almost lattice matched to GaN [1, 2, 5-7]. The large band offset between these two materials predicted using density functional theory (DFT) calculations has been the enabling factor for the calculated remarkable efficiency improvement in recently proposed III-N/ZnGeN$_2$ heterostructure-based light-emitting diodes (LEDs) [8], quantum cascade lasers [9] and UV LEDs [10].

To date only calculations of the valence band offset (VBO) between the Pna2$_1$ phase of pure ZnGeN$_2$ and GaN have been reported. Punya et al. [11, 12] used DFT employing the local density approximation (LDA) for the electrostatic potential profile, and quasiparticle self-consistent GW calculations (where G is the one-electron Green's function and W the screened Coulomb interaction) for the band edge positions relative to the average electrostatic potential in each material. They included the strain effects in ZnGeN$_2$ to match the in-plane lattice constant to the unstrained GaN substrate. They found that the valence band maximum (VBM) of ZnGeN$_2$ lies well above that of GaN [11, 12]: the predicted VBO at the ZnGeN$_2$/GaN heterointerface with the normals along the Pbn2$_1$ [100], [010] and [001] directions were reported to be 1.44 eV, 1.36 eV, and 1.38 eV, respectively [12]. On the other hand, the "natural" valence band offset was determined from the calculated electron affinity values to be 0.49 eV for the (100) orientation of Pbn2$_1$ ZnGeN$_2$ [13]. The effect of the Zn 3$d$ bands on the VBM can explain the positive VBO, with the valence band of ZnGeN$_2$ above that of GaN [11, 14]. The Zn 3$d$ bands have stronger hybridization than the Ga 3$d$ bands since they lie significantly closer to the VBM as compared to the Ga 3$d$ bands. This situation will cause the VBM in ZnGeN$_2$ to move to a higher energy as compared to that in GaN [11, 14].

Recently, a DFT-based calculation using a hybrid functional [15] for surfaces combined with the electron-affinity rule found the natural VBM of ZnGeN$_2$ to be 0.28 eV below that of GaN, a result that is strikingly different from the results of Ref. [11] and [12]. The origin of this discrepancy is presently not clear.

Compared to GaN, the development of ZnGeN$_2$ is still at a very early stage. There have been only a handful of reports on the growth of ZnGeN$_2$ thin films. These include halide vapor phase epitaxy growth on sapphire [6], metalorganic chemical vapor deposition (MOCVD) growth on sapphire with (0001) c- [7, 16], (11-20) r- [5, 7, 16], and (10-12) a- [16] out-of-plane orientations as well as on GaN/c-sapphire templates [17], and molecular beam epitaxy (MBE) growth on GaN/c-sapphire templates [18]. MOCVD growth of (ZnGe)$_{1-x}$Ga$_{2x}$N$_2$ thin films on c- and r-plane sapphire as well as GaN/c-sapphire was also reported [19]. Experimental investigations of ZnGeN$_2$ have mainly focused on the morphological [5, 16-19], optical [3, 5, 16, 17, 20], crystal structural [5, 7, 16-21], and lattice vibrational properties [21-23]. Deep-level defects in MOCVD-grown ZnGeN$_2$ films on sapphire substrates have also been investigated [24]. Here we report on measurements of the valence band offsets of (ZnGe)$_{1-x}$Ga$_{2x}$N$_2$ with GaN for x = 0 (ZnGeN$_2$) and x = 0.06 ((ZnGe)$_{0.94}$Ga$_{0.12}$N$_2$) at the respective heterointerfaces using X-ray photoemission spectroscopy (XPS).

## 2. Experimental details

## 2.1. MOCVD growth

For this study, two sets of $(ZnGe)_{1-x}Ga_{2x}N_2$ and $(ZnGe)_{1-x}Ga_{2x}N_2$/GaN samples were grown on commercially purchased GaN/c-sapphire templates in a custom-designed dual chamber vertical-flow, rotating-disc MOCVD system having a showerhead gas injection configuration. Diethylzinc (DEZn), germane ($GeH_4$) and ammonia ($NH_3$) were used as the precursors for Zn, Ge, and N, respectively. Nitrogen ($N_2$) was used as the carrier gas. The total reactor pressure was set at 500 Torr. The DEZn/GeH_4 molar flow rate was optimized for obtaining single crystalline thin films with stoichiometric cation composition (Zn/(Zn+Ge) = 0.5) and smooth surface morphology. The detailed growth technique was reported in Ref. 17. Two $ZnGeN_2$ samples were grown on GaN templates at growth temperature $T_G$ = 650 °C, for 10 s and 80 s, using the same growth conditions for both. Two $ZnGeN_2$-GaN alloy samples were grown at $T_G$ = 735 °C for 23 s and 165 s on GaN templates. An in-situ reflectometer was used to estimate the thicknesses of the films. Growth rates were also confirmed from cross-sectional scanning electron microscopy (SEM) images of thicker films. The thicknesses of the $ZnGeN_2$ samples were ~2 nm and ~16 nm for the growth durations of 10 s and 80 s, respectively whereas the thicknesses of the $ZnGeN_2$-GaN alloy samples were ~3 nm and ~20 nm for the growth durations of 23 s and 165 s, respectively.

## 2.2. Materials characterization

The quality of the heterointerfaces was investigated using high magnification scanning transmission electron microscopy (STEM) images captured by a Thermofisher probe-corrected Titan STEM operated at 300 kV. SEM imaging were carried out by a Helios Nanolab 600 and an FEI Apreo LoVac analytical SEM. Surface roughness of the films were determined from atomic force microscopy (AFM) images using a Bruker Icon 3 AFM. X-ray diffraction (XRD) measurements were carried out using a Bruker D8 Discover XRD with the Cu Kα source. The XPS measurements on the $ZnGeN_2$ samples were carried out using a PHI 5000 VersaProbeTM system equipped with a Scanning XPS Microprobe X-ray source with hυ (Al Kα) = 1486.6 eV, FWHM≤ 0.5 eV. The XPS measurements on the $ZnGeN_2$-GaN alloy samples were performed using a Kratos Axis Ultra X-ray photoelectron spectrometer using a monochromatic Al Kα X-ray source, $E_{photon}$ = 1486.6 eV. The measurements were performed on as-grown samples without performing additional surface cleaning before or after loading into the XPS chambers to prevent potential surface contamination or structural damage [25]. The samples were stored in the chamber for 18 hours under high vacuum prior to the XPS measurements.

## 3.   Results and discussions

Figure 1 shows schematic diagrams of the samples. Figures 2(a) and 2(b) present the high magnification STEM images of a $ZnGeN_2$ and a $(ZnGe)_{0.94}Ga_{0.12}N_2$ film grown on GaN grown under identical conditions to the corresponding samples used to measure the band offsets in this work. The red arrows mark the interfaces which show clear contrasts between the substrate and the films. The STEM images demonstrate the high-quality interfaces between the binary GaN and ternary $ZnGeN_2$ or quaternary $ZnGeN_2$-GaN alloy lattices. To investigate the surface morphologies, FESEM images and AFM images were obtained from a ~350 nm thick $ZnGeN_2$ film and a ~1 μm thick $ZnGeN_2$-GaN alloy film, which were grown using the identical growth conditions as those described above. Plan-view SEM images in figures 3 (a), and 3 (b) show planar surfaces for both films. The RMS roughness values obtained from 5 μm × 5 μm AFM images (figures 3 (c), and 3 (d)) were 3.0 nm and 1.9 nm, respectively. Based on the XRD 2θ-ω scan profiles (not shown), $ZnGeN_2$ and $(ZnGe)_{1-x}Ga_{2x}N_2$ thin films grown under various similar conditions to those used in this work were single crystalline [17, 19]. For $ZnGeN_2$ films grown on GaN, very close XRD 2θ positions of the film and the substrate (e.g., for ((002) peaks Δ2θ ~ 0.15°) [17]) makes it challenging to separate the signals between the two from XRD ω-rocking curves. For ZnGeN₂ films grown on c-sapphire substrates, a *full width at half maxima* (FWHM) of the ω-rocking curve around the (002) peak as low as 0.28° was obtained (figure 3 (e)). Crystalline quality of the ZnGeN₂ films grown on GaN (lattice mismatch < 1%) is significantly improved as compared to the ZnGeN₂ films grown on c-sapphire (~16% lattice mismatch). A detailed investigation of crystal structural and surface morphological properties of ZnGeN₂ films was reported in Ref. 17.

The atomic compositions of Zn, Ge, and Ga in the samples were determined using the respective $2p$ XPS peaks. In the ZnGeN₂ as well as the ZnGeN₂-GaN alloy samples, the Zn/(Zn+Ge) compositions were close to the stoichiometric value (0.50±0.07). The Ga/(Zn+Ge+Ga) compositions in the 20 nm thick quaternary alloy sample was 0.12±0.02, which corresponds to x=0.06 in $(ZnGe)_{1-x}Ga_{2x}N_2$. Hereafter, the alloy samples will be referred to as $(ZnGe)_{0.94}Ga_{0.12}N_2$. The out-of-plane orientation of $(ZnGe)_{1-x}Ga_{2x}N_2$ grown on GaN/c-sapphire is (001). Therefore, the measured VBO values correspond to a (001) $(ZnGe)_{1-x}Ga_{2x}N_2$/(001) GaN heterointerface.

The valence band offset, ΔEᵥ, between two materials A and B can be determined by using Kraut's method [26] according to which

$$\Delta E_v = \left( E_{CL,b}^B - E_v^B \right) - \left( E_{CL,b}^A - E_v^A \right) - \left( E_{CL,i}^B - E_{CL,i}^A \right) \qquad (1)$$

where, $E_{CL,b}^{A/B}$ is the binding energy of the core level in the bulk material A/B, $E_{CL,i}^{A/B}$ is the binding energy of the core level of the material A/B determined from the heterostructure between A and B and $E_V^{A/B}$ is the position of the VBM in the bulk material A/B. The positions of the core levels are determined with respect to the Fermi level. A positive value of $\Delta E_v$ would indicate the VBM of B to be above that of A.

XPS is a commonly used technique to determine the parameters in Eq. (1) [27, 29, 30]. The parameters in the first two terms in Eq. (1) are determined using XPS measurements of the bulk samples of materials A and B. The parameters in the third term are determined using XPS measurement of a heterostructure formed between the two materials. In order to be able to measure the core level of the bottom layer of the two-layer heterostructure using XPS, the photoelectrons from this layer need to be detected. A thick top layer can absorb all of the photoelectrons emitted from the bottom layer due to the finite escape depth of the photoelectrons [25, 27, 28], as illustrated in figure 1. Therefore, the top layer needs to be sufficiently thin so that a reasonable number of photoelectrons from the bottom layer can escape and be detected.

With a view to determining the band offset between $ZnGeN_2$ and GaN at a $ZnGeN_2$/GaN heterointerface, XPS measurements were performed on the GaN template, the 16-nm-thick $ZnGeN_2$ layer and the 2-nm-thick $ZnGeN_2$/GaN heterostructure samples. The binding energies of the adventitious C $1s$ peaks were measured to be 283.9 eV, 286.1 eV, and 284.9 eV, respectively. Figures 4(a-c) present the XPS spectra near the Ge $2p$, Ga $2p$ and Zn $2p$ core-level positions, respectively, for all three samples. As can be seen here, the Ga $2p$ peak is absent in the spectra from the 16-nm-thick $ZnGeN_2$ sample, which indicates that this sample has sufficient thickness to be characteristic of bulk $ZnGeN_2$. In addition, only the Ga $2p$ peak was observed in the XPS spectra of the GaN template. The XPS spectra of the 2-nm-thick $ZnGeN_2$/GaN sample showed Ge $2p$ and Zn $2p$ as well as Ga $2p$ peaks. Therefore, this sample is characteristic of $ZnGeN_2$/GaN heterostructures. The XPS spectra near the Ge $3d$, Ga $3d$, and Zn $3d$ bands obtained from the GaN, 16-nm-thick $ZnGeN_2$ and 2-nm-thick $ZnGeN_2$/GaN heterostructure samples are shown in figures 4(d), 4(e), and 4(f), respectively, by the black circles. The blue dashed lines were fitted to the peaks by assuming a Voigt line shape and Shirley background. The Shirley backgrounds are shown by the green dash-dot curves. The Zn $3d$ peaks overlapped with the N $2p$ peaks for both the 16-nm-thick ZnGeN$_2$ and the 2-nm-thick ZnGeN$_2$/GaN heterostructure samples, shown in in figures 4(e), and 4(f), respectively. For the GaN sample (magenta spectra), only the Ga $3d$ peak, at a binding energy of 18.6 eV, was observed. On the other hand, for the ZnGeN$_2$ sample (blue spectra), only the Ge $3d$ and Zn $3d$ peaks were observed, at binding energies 33.0 eV and 11.4 eV, respectively. All three (Ge, Ga, and Zn) $3d$ peaks were observed in the spectra from the 2 nm ZnGeN$_2$/GaN heterostructure sample (red spectra), at binding energies 31.9 eV, 20.25 eV and 10.5 eV, respectively. The FWHM values were 1.0 eV for Ga $3d$ peaks in figure 4(d) and figure 4(f), 1.7 eV for Ge $3d$ peaks and 1.6-1.7 eV for Zn $3d$ peaks. Please note that no corrections in the peak positions were done since only the differences between the spectral features from any given sample are used to calculate the VBO, namely, core level differences between GaN substrate and ZnGeN$_2$ overlayer on the one hand and VBM vs. the same core levels from separate bulk-like samples (GaN substrate and thick ZnGeN$_2$ sample) on the other hand. Any offset in absolute binding energies between different samples in Eq. 1 will be cancelled out. This is precisely the advantage of the Kraut method. The measured binding energies for the core levels are within the range of corresponding values reported previously [29].

As noted earlier, the binding energies of the Zn and Ge core levels obtained from the ZnGeN$_2$ sample give $E_{CL,b}^{ZnGeN_2}$, the binding energies of the Ga core levels obtained from the GaN sample give $E_{CL,b}^{GaN}$. The corresponding values from the ZnGeN$_2$/GaN heterostructure sample give $E_{CL,i}^{ZnGeN_2}$ or $E_{CL,i}^{GaN}$. The VBM values, $E_V^{GaN}$ and $E_V^{ZnGeN_2}$, were obtained from the GaN and ZnGeN$_2$ samples, respectively. Figure 4(g) presents the XPS spectra near the valence band edge of the 16 nm ZnGeN$_2$ sample and the GaN template. The VBM values were calculated from the intersection of the straight line fitted over the leading edge of the XPS spectra with the average of the flat portion of the spectra for energies above the valence band maxima. Table 1 lists the values of the VBM as well as the core level binding energies. The calculated VBO between ZnGeN$_2$ and GaN from Eq. (1) is 1.45±0.15 eV when Zn $3d$ and Ga $3d$ core-level energies are used and 1.65±0.15 eV when Ge $3d$ and Ga $3d$ core-level energies are used with the average being 1.55±0.15 eV. The calculated VBO value using Zn $2p_{3/2}$ and Ga $2p_{3/2}$, and Ge $2p_{3/2}$ and Ga $2p_{3/2}$ core level energies is 1.54±0.15 eV, and 1.74±0.15 eV, respectively, which are consistent with the values calculated using corresponding $3d$ core-level energies. These VBO values are in reasonable agreement with those obtained from explicit interface calculations (1.4 eV) [11, 12] rather than the electron-affinity based values. [13, 15].

The same approach was used to determine the VBO of (ZnGe)$_{0.94}$Ga$_{0.12}$N$_2$ with GaN. The XPS spectra collected from (ZnGe)$_{0.94}$Ga$_{0.12}$N$_2$ samples are shown in figure 5. The spectra measured from a bare piece of GaN template on which the (ZnGe)$_{0.94}$Ga$_{0.12}$N$_2$ samples were grown are also shown. The binding energies of the adventitious C $1s$ peaks were measured to be 286.1 eV, 285.8 eV, and 285.6 eV, respectively. The Ge $2p$ and Zn $2p$ peaks were observed in figure 5(a) and figure 5(c), respectively, from both the 3 nm and 20 nm thick (ZnGe)$_{0.94}$Ga$_{0.12}$N$_2$ samples, but not from the GaN template. The Ga $2p$ peak was observed in all three samples. The 20 nm (ZnGe)$_{0.94}$Ga$_{0.12}$N$_2$ is sufficiently thick to absorb all of the XPS signals originating in the underlying GaN substrate. Therefore, the Ga $2p$ peaks in the spectra from the 20-nm-thick (ZnGe)$_{0.94}$Ga$_{0.12}$N$_2$ sample, shown in blue, can be assigned to the Ga atoms in the top quaternary alloy. Figures 5(d-f) show the XPS spectra near the Ge $3d$, Ga $3d$, and Zn $3d$ bands obtained from the GaN, 20-nm-thick (ZnGe)$_{0.94}$Ga$_{0.12}$N$_2$ and 3-nm-thick (ZnGe)$_{0.94}$Ga$_{0.12}$N$_2$/GaN heterojunction samples, respectively, in black circles. The dashed blue curves in figures 5(d-f) are the fitted peaks assuming Voigt peak shapes and Shirley background. The overlapped Zn $3d$ and N $2p$ peaks are marked in figures 5(e) and 5(f). In the case of the 20-nm-thick $(ZnGe)_{0.94}Ga_{0.12}N_2$ sample in figure 5(e) two components were necessary to fit the Ge $3d$ peak – The peak at the lower binding energy probably corresponds to a lower charge state of Ge present in the sample [30]. The binding energies of the Ge $3d$ and Zn $3d$ peaks were determined to be 32.7 and 11.3 eV, respectively, for the 20-nm-thick $(ZnGe)_{0.94}Ga_{0.12}N_2$ (bulk) sample, and 32.6 eV and 11.2 eV, respectively, for the 3-nm-thick $(ZnGe)_{0.94}Ga_{0.12}N_2$ (heterostructure) sample. The FWHM values were 1.4 eV for Ge $3d$ peaks and 1.4-1.6 eV for Zn $3d$ peaks. The binding energy of the Ga $3d$ peak was 20.9 eV in both the bulk GaN and the 3-nm-thick $(ZnGe)_{0.94}Ga_{0.12}N_2$ heterostructure sample. The FWHM of these Ga $3d$ peaks are 1.0 eV. The positions of the valence band maxima in the 20-nm-thick $(ZnGe)_{0.94}Ga_{0.12}N_2$ (bulk) sample and the GaN template are marked in figure 5(g). These are at 2.49±0.02 eV and 3.68±0.05 eV, respectively. The difference (2.58 eV) between the positions of the VBM of GaN templates in figure 4(g) and figure 5(g) is comparable to the difference (2.2 eV) between the binding energies of the adventitious C $1s$ levels of the two samples, which is probably caused by the difference in electrical conductivity of the two templates. The difference in electrical conductivity can result from different concentrations of unintentional impurities in the film, for example, C, which in turn can cause shifts in the core-level positions [31]. The core energy levels and the positions of the VBM determined for the $(ZnGe)_{0.94}Ga_{0.12}N_2$ samples are listed in Table 2. The VBO between $(ZnGe)_{0.94}Ga_{0.12}N_2$ and GaN calculated using either Zn or Ge $3d$ core levels is 1.29±0.2 eV, which is 0.26 eV lower than the average VBO determined for $ZnGeN_2$ using the $3d$ core-level energies. This lowering of the VBO is attributed to the reduced effect of the Zn $3d$ bands in the $ZnGeN_2$-GaN alloy compared to in $ZnGeN_2$. Assuming a linear dependence of the VBO with composition x of $(ZnGe)_{1-x}Ga_{2x}N_2$, the predicted VBO of $(ZnGe)_{0.94}Ga_{0.12}N_2$ from theoretically calculated VBO of $ZnGeN_2$ [12] would be 1.32 eV, which is very close to the experimentally determined value. The error bars in the determined VBO values are mainly due to the uncertainties in the determined VBM values. The calculated total polarization difference at the $ZnGeN_2/GaN$ heterointerface is very small [15, 32], therefore, errors in the determined VBO values due to polarization induced band bending is expected to be small. For instance, the $ZnGeN_2/GaN$ heterostructure VBOs, predicted by Punya et al. [12], along the polar and the non-polar directions differ by < 0.1 eV.

The conduction band offsets were calculated using the determined VBO values and the band gap of the materials using the relation $\Delta E_C = \Delta E_V + (E_g^{(ZnGe)_{1-x}(Ga)_{2x}N_2} - E_g^{GaN})$. The band gaps of $ZnGeN_2$ and $(ZnGe)_{0.94}Ga_{0.12}N_2$ were estimated by analyzing the inelastic energy loss features of N 1s peaks in the XPS spectra of 16-nm-thick $ZnGeN_2$ and 20-nm-thick $(ZnGe)_{0.94}Ga_{0.12}N_2$ samples. Before reaching the XPS detector, a photo-emitted electron can lose energy to (i) high frequency plasma oscillations in the valence band, (ii) surface plasmons and (iii) another electron transitioning from the valence band into the conduction band. The band-to-band transition involves the minimum-energy process among the three, and therefore, the band gap of the material determines the lower limit of the energy loss of the photo-emitted electrons due to inelastic processes [33]. Energy loss features of the N 1$s$ peak in XPS spectra were previously used to determine the bandgap of $Si_3N_4$ [34]. Figures 6(a) and 6(b) show the XPS spectra near the N 1$s$ core levels obtained from the 16-nm-thick $ZnGeN_2$ and 20-nm-thick $(ZnGe)_{0.94}Ga_{0.12}N_2$ samples, respectively. The approximate onsets of the inelastic energy losses were estimated by linear fitting to the energy-loss features on the higher binding energy side of the N 1$s$ peak. The difference between the binding energies corresponding to the N 1$s$ peak and the onset of the inelastic energy loss provides the bandgap of the epilayer. The band gaps found by this method are 3.0±0.2 eV for ZnGeN$_2$ and 3.1±0.2 eV for (ZnGe)$_{0.94}$Ga$_{0.12}$N$_2$. The error bars correspond to the standard deviation of the estimated onset of the inelastic energy losses. The slight increase in the band gap of the 50:50 alloy of ZnGeN$_2$-GaN alloy ((ZnGe)$_{0.5}$GaN$_2$) as compared to those of ZnGeN$_2$ or GaN was predicted by first-principles calculations [14]. The lower values of the band gaps as compared to the predicted values are probably due to the presence of disorder in the cation sublattice [35]. The band alignments of ZnGeN$_2$ and (ZnGe)$_{0.94}$Ga$_{0.12}$N$_2$ with GaN are shown in figure 6(c) using the average of the VBO values determined using the Zn 3$d$ and Ge 3$d$ bands.

## 4. Conclusions

In conclusion, the valence band offsets of MOCVD-grown (ZnGe)$_{1-x}$Ga$_{2x}$N$_2$ with GaN, with x = 0 and 0.06, measured using XPS, are presented for the first time to the best of our knowledge. The measured VBO for ZnGeN$_2$ (1.45 eV - 1.65 eV) are comparable to the predicted value from first-principles calculations using explicit interface calculations [11, 12]. For (ZnGe)$_{0.94}$Ga$_{0.12}$N$_2$, the VBO was measured to be 1.29 eV, which is very close to the predicted value from theoretically calculated VBO of ZnGeN$_2$ assuming a linear dependence of VBO on composition. The results from this study will expand device designs based on pure III-nitrides to III-nitrides/II-IV-N$_2$, which can potentially address key challenges in III-nitride based electronic and optoelectronic device technologies.

**Acknowledgements**

The authors acknowledge funding support from the U.S. Department of Energy (DE-EE0008718) and from the National Science Foundation (DMREF-SusChEM-1533957). Karim and Zhao also acknowledge support from the Seed Grant from the Institute for Materials Research at the Ohio

State University and the Center for Emergent Materials, an NSF-funded MRSEC under award DMR-1420451. Brillson and Noesges also acknowledge support from National Science Foundation grant DMR 18-00130 (Tanya Paskova).

## Data Availability

The data that support the findings of this study are available from the corresponding author upon reasonable request.

**Table 1**

**Table 1.** Positions of the bulk valence band maxima and the core Ga, Zn and Ge levels extracted from the XPS spectra, from the GaN template, the 16 nm ZnGeN$_2$ (bulk) and the 2 nm ZnGeN$_2$/GaN (heterostructure) samples. The binding energies of the core levels were determined by peak fitting assuming a Voigt shape and Shirley background. No correction in the peak positions was done. The binding energies of adventitious C 1$s$ peaks are listed for reference.

| Sample | C | Ga | | Zn | | Ge | | E$_V$ |
|---|---|---|---|---|---|---|---|---|
| | E$_{1s}$ (eV) | E$_{Ga\text{-}2p3/2}$ (eV) | E$_{Ga\text{-}3d}$ (eV) | E$_{Zn\text{-}2p3/2}$ (eV) | E$_{Zn\text{-}3d}$ (eV) | E$_{Ge\text{-}2p3/2}$ (eV) | E$_{Ge\text{-}3d}$ (eV) | VBM (eV) |
| GaN | 283.9 | 1116.4 | 18.6 | - | - | - | - | 1.1 ±0.07 |
| ZnGeN$_2$ | 286.1 | - | - | 1023.1 | 11.4 | 1221.0 | 33.0 | 2.26±0.09 |
| ZnGeN$_2$/GaN | 284.9 | 1118.0 | 20.3 | 1022.0 | 10.5 | 1219.7 | 31.9 | - |

**Table 2.** Position of the bulk valence band maxima and the core Ga, Zn and Ge levels extracted from the XPS spectra of the GaN template, the 20 nm $(ZnGe)_{0.94}Ga_{0.12}N_2$ (bulk) and the 3 nm $(ZnGe)_{0.94}Ga_{0.12}N_2/GaN$ (heterostructure) samples. The binding energies of the core levels were determined by peak fitting assuming a Voigt shape and Shirley background. No correction in the peak positions was done. The binding energies of adventitious C $1s$ peaks are listed for reference.

| Sample | C | Ga | | Zn | | Ge | | $E_V$ |
|---|---|---|---|---|---|---|---|---|
| | $E_{1s}$ (eV) | $E_{Ga-2p3/2}$ (eV) | $E_{Ga-3d}$ (eV) | $E_{Zn-2p3/2}$ (eV) | $E_{Zn-3d}$ (eV) | $E_{Ge-2p3/2}$ (eV) | $E_{Ge-3d}$ (eV) | VBM (eV) |
| GaN | 286.1 | 1118.7 | 20.9 | - | - | - | - | 3.68±0.05 |
| $(ZnGe)_{0.94}Ga_{0.12}N_2$ | 285.8 | 1118.8 | 21.0 | 1022.7 | 11.3 | 1220.5 | 32.7 | 2.49±0.02 |
| $(ZnGe)_{0.94}Ga_{0.12}N_2/GaN$ | 285.6 | 1118.6 | 20.9 | 1022.6 | 11.2 | 1220.3 | 32.6 | - |

**Figure 1**

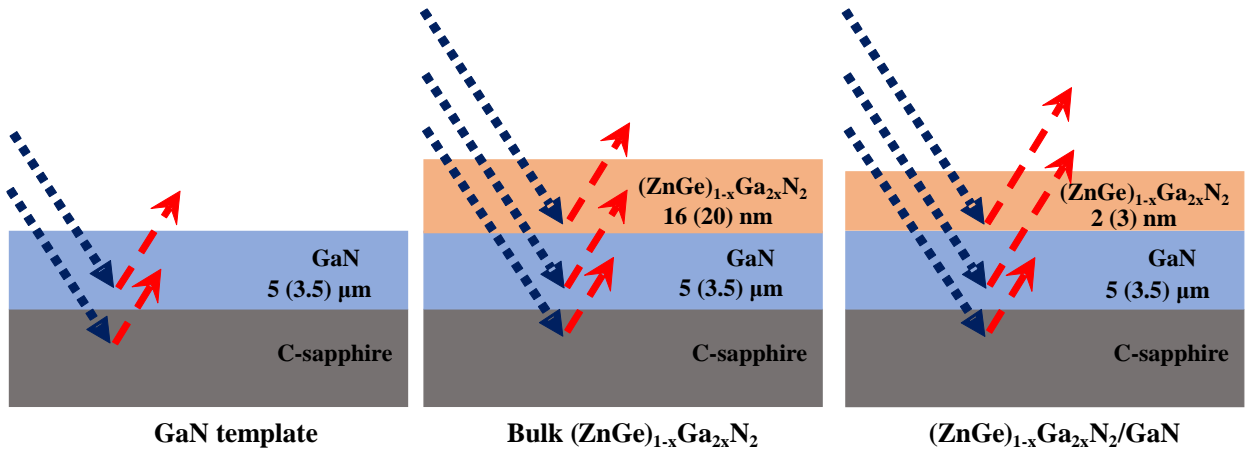

**Figure 1.** Schematics of the bulk GaN, bulk $(ZnGe)_{1-x}Ga_{2x}N_2$ and $(ZnGe)_{1-x}Ga_{2x}N_2/GaN$ heterostructure samples used to determine the valence band offset of $ZnGeN_2/GaN$ and $(ZnGe)_{0.94}Ga_{0.12}N_2/GaN$ using XPS. The thicknesses are 5 μm (3.5 μm), 16 nm (20 nm), and 2 nm (3 nm), respectively, for $ZnGeN_2/GaN$ ($(ZnGe)_{0.94}Ga_{0.12}N_2/GaN$). The blue dotted lines indicate the incident X-ray beam and red dashed lines indicate the emitted photoelectrons. The arrows indicate the directions of propagation. Short red arrows indicate that the photoelectrons are absorbed before escaping from the sample.

**Figure 2**

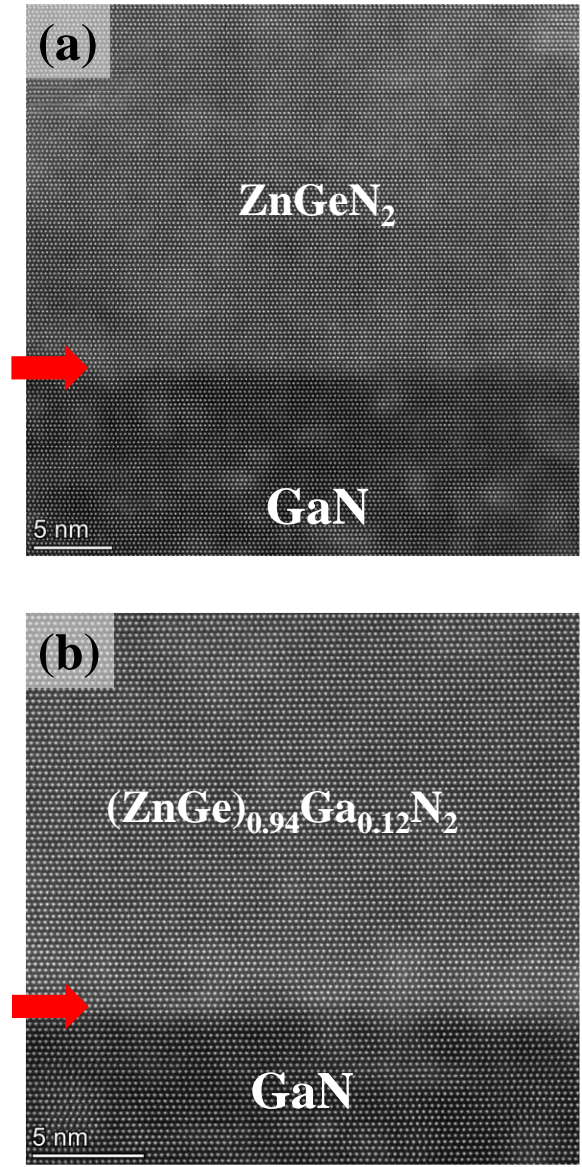

**Figure 2.** High magnification STEM images showing the interfaces of (a) a ZnGeN$_2$ film and (b) a ZnGeN$_2$-GaN alloy film on GaN. The interface is marked by the red arrows.

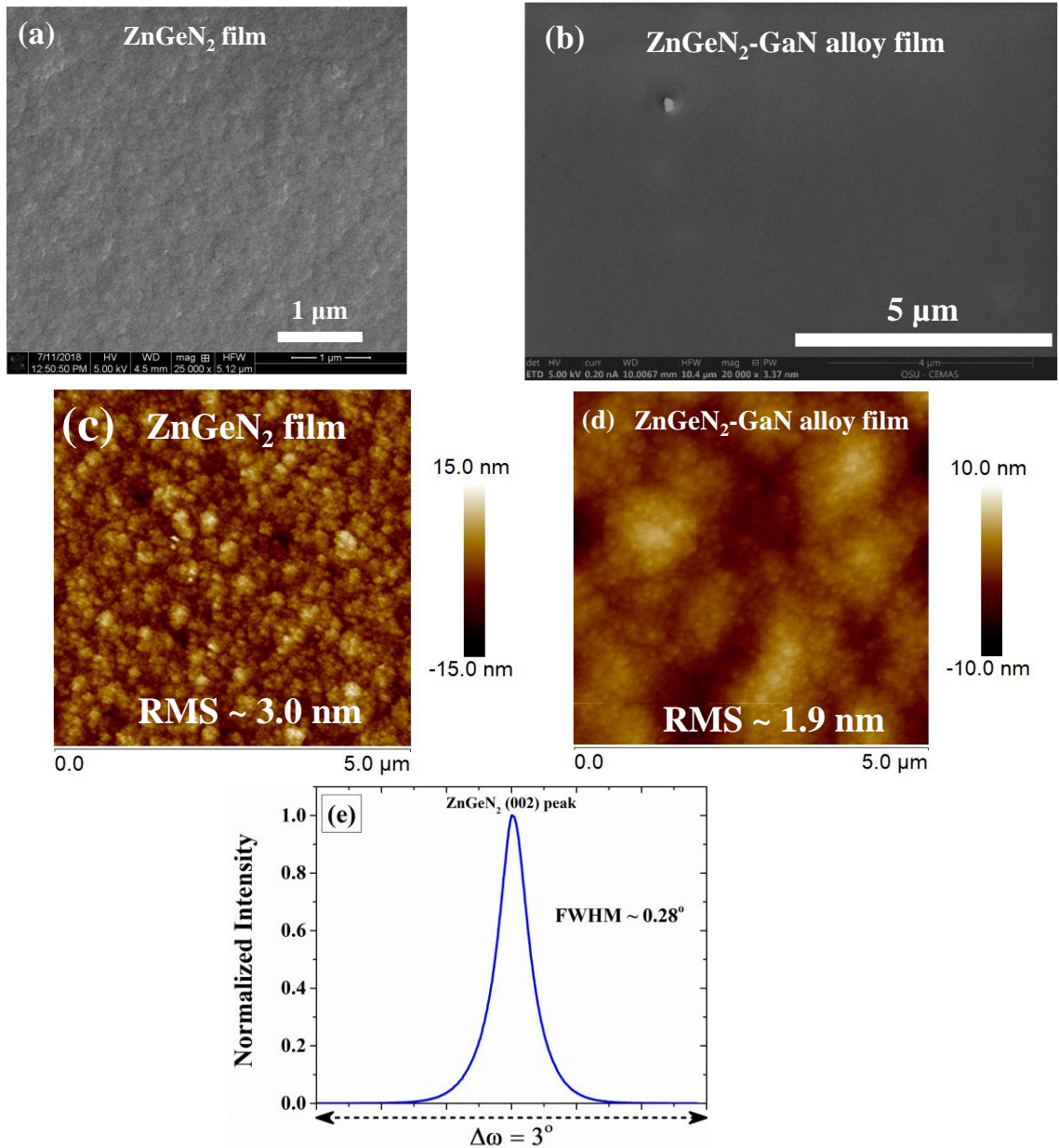

**Figure 3**. (a, b) Plan-view FESEM images and (c, d) 5 μm × 5 μm AFM images of a ~350 nm thick ZnGeN₂ film (a, c) and a ~1 μm thick ZnGeN₂-GaN alloy film (b, d) grown using identical growth conditions as those used for determining the valence band offset in this work. (e) ω-rocking curve around the ZnGeN₂ (002) peak obtained from a single crystalline ZnGeN₂ film grown on c-sapphire.

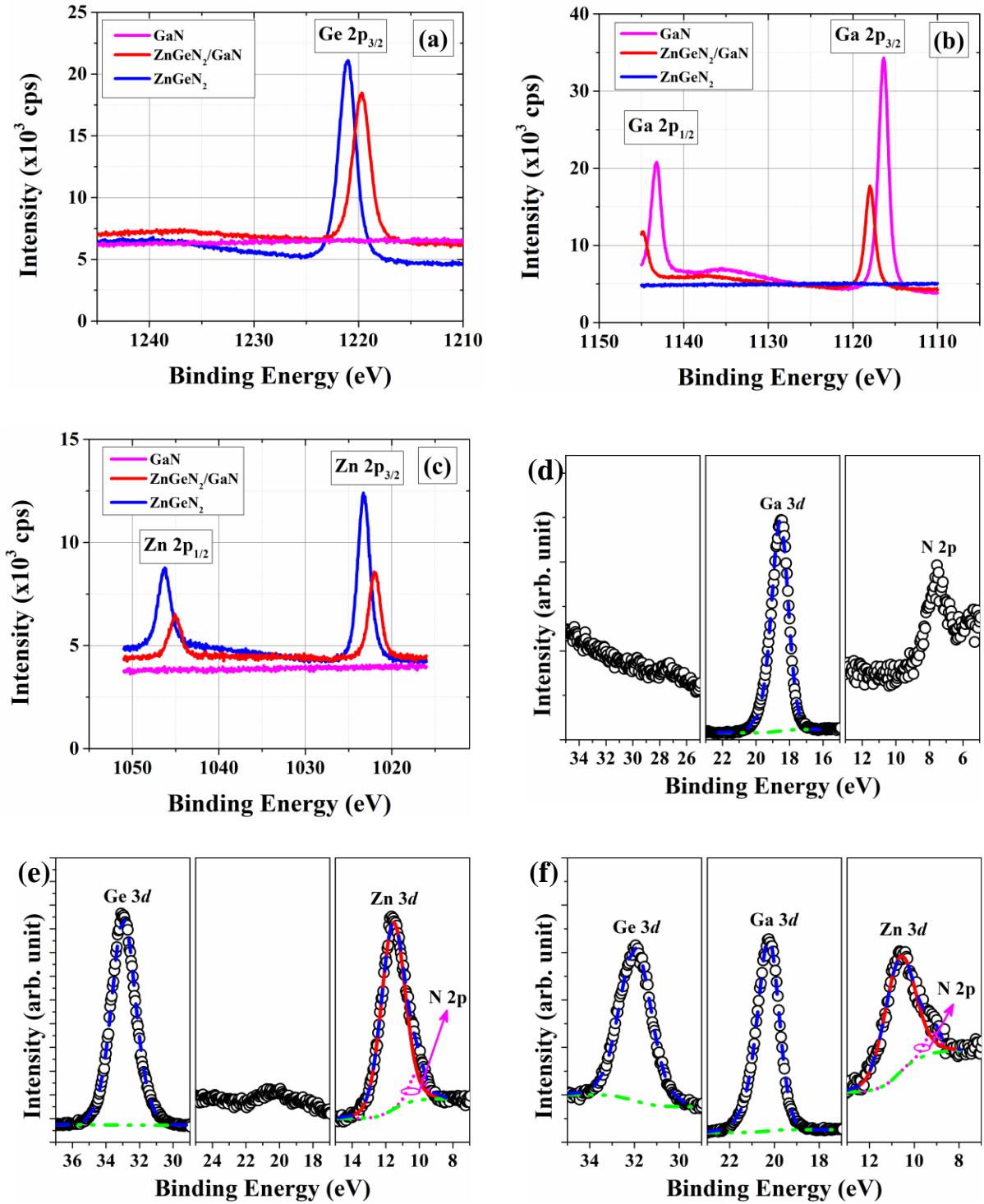

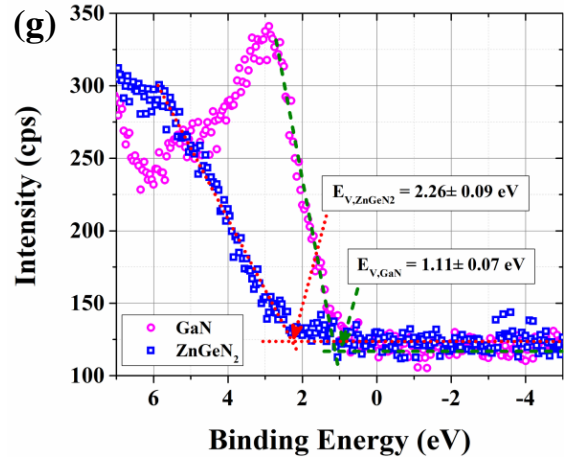

**Figure 4.** The XPS spectra near the (a) Ge 2*p* region, (b) *Ga* 2*p* region and (c) Zn 2*p* region obtained from the GaN, 16 nm ZnGeN₂ and the 2 nm ZnGeN₂/GaN heterostructure samples. (d-f) XPS spectra near the Ge 3*d*, Ga 3*d*, and Zn 3*d* bands obtained from the GaN, 16 nm ZnGeN₂ and the 2 nm ZnGeN₂/GaN heterostructure samples, respectively. The original spectra are shown by the black circles. The dashed blue curves are the fitted peaks obtained assuming a Voigt shape and Shirley background. The background is shown by the green dash-dot lines. The Zn 3*d* peaks overlapped with the N 2*p* peaks which are marked in respective panels. The full width at half maxima of the fitted peaks 1.7 eV, 1.0 eV and 1.6-1.7 eV for Ge 3*d*, Ga 3*d* and Zn 3*d* bands, respectively. (g) The XPS spectra showing the valence band edges of the GaN (magenta circles) and the 16 nm ZnGeN₂ (blue squares) samples.

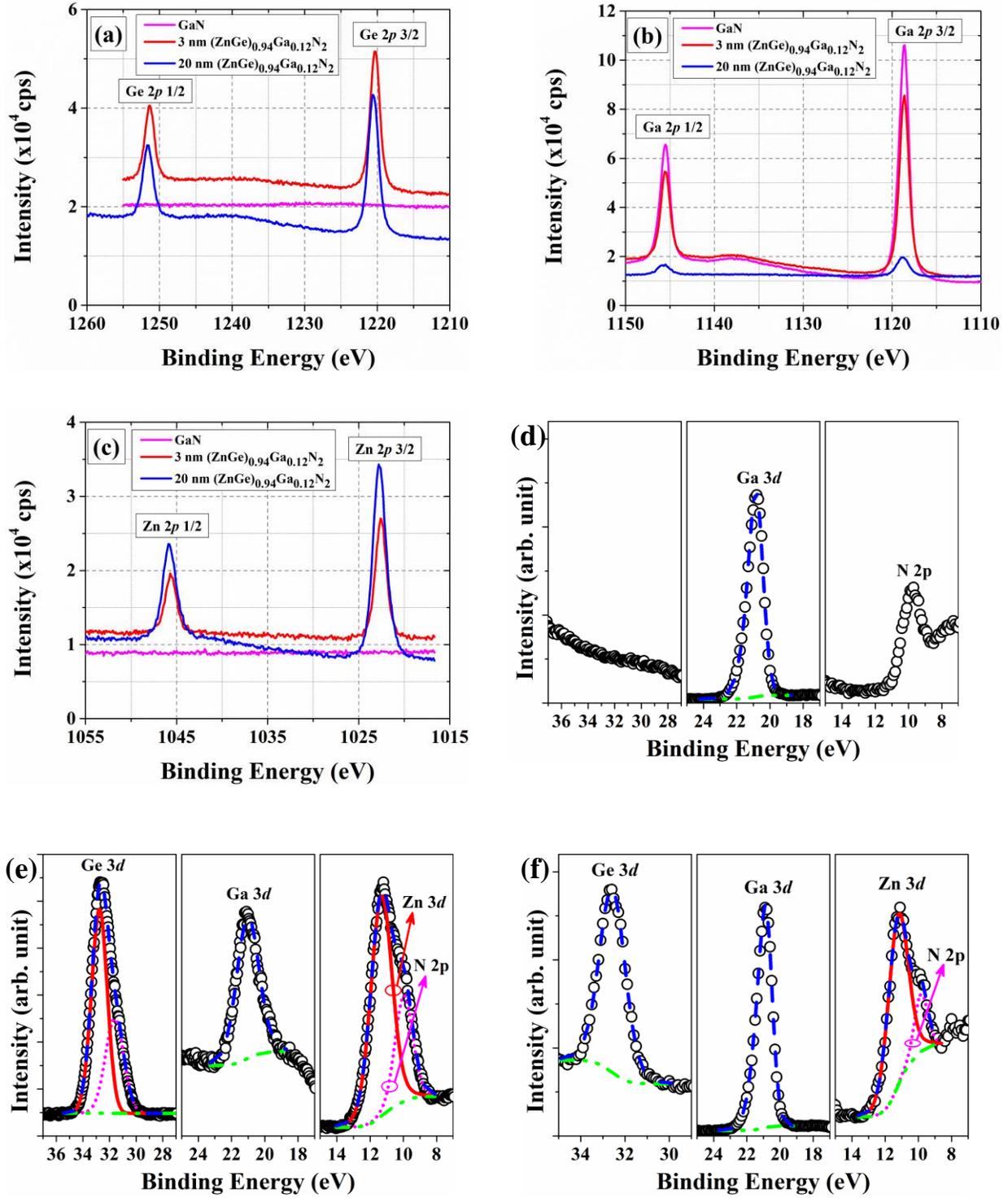

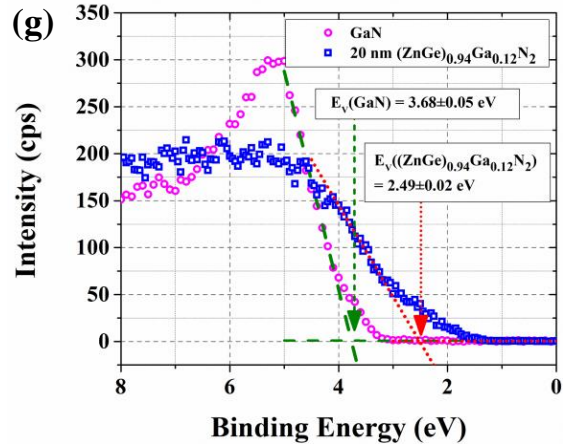

**Figure 5.** The XPS spectra near the (a) Ge 2*p* region, (b) *Ga* 2*p* region and (c) Zn 2*p* region obtained from the GaN, 20 nm $(ZnGe)_{0.94}Ga_{0.12}N_2$ and the 3 nm $(ZnGe)_{0.94}Ga_{0.12}N_2$ /GaN heterostructure samples. (d-f) XPS spectra near the Ge 3*d*, Ga 3*d*, and Zn 3*d* bands obtained from the GaN, 20 nm $(ZnGe)_{0.94}Ga_{0.12}N_2$ and the 3 nm $(ZnGe)_{0.94}Ga_{0.12}N_2$/GaN heterostructure samples, respectively. The original spectra are shown by the black circles. The dashed blue curves are the fitted peaks obtained assuming a Voigt shape and Shirley background. The background is shown by the green dash-dot lines. The Zn 3*d* peaks overlapped with the N 2*p* peaks which are marked in respective panels. For the 20 nm $(ZnGe)_{0.94}Ga_{0.12}N_2$ sample, two components were required to fit the Ge 3*d* peak. The full width at half maxima of the fitted peaks 1.4 eV, 1.0 eV and 1.4-1.5 eV for Ge 3*d*, Ga 3*d* and Zn 3*d* bands, respectively. (g) The XPS spectra showing the valence band edges of the GaN (magenta circles) and the 20 nm $(ZnGe)_{0.94}Ga_{0.12}N_2$ (blue squares) sample.

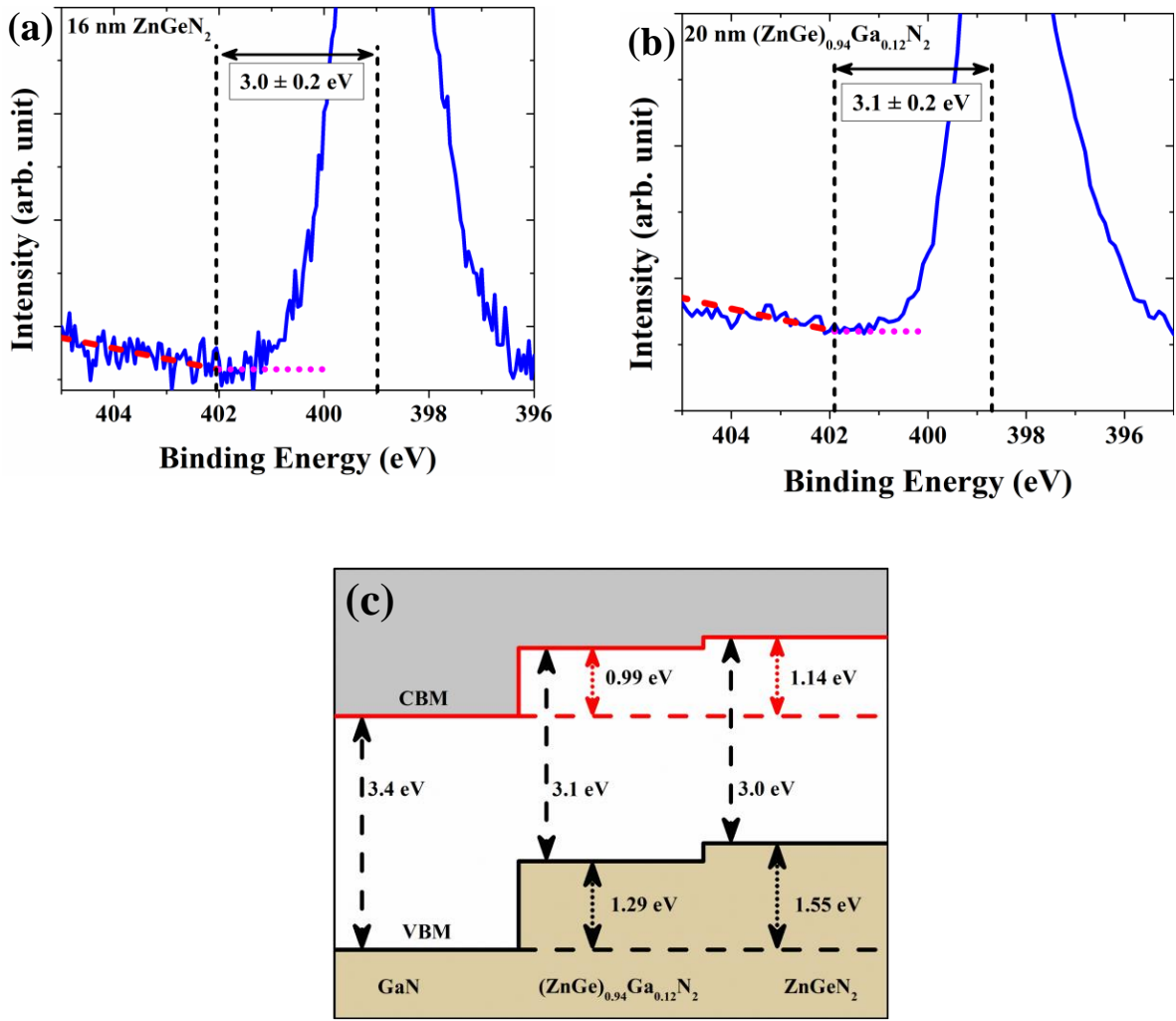

**Figure 6.** The XPS spectra showing the N $1s$ peak and onset of inelastic energy loss obtained from (a) 16-nm-thick ZnGeN$_2$ and (b) 20 nm (ZnGe)$_{0.94}$Ga$_{0.12}$N$_2$ samples. (c) Band alignments of (ZnGe)$_{1-x}$Ga$_{2x}$N$_2$ (x = 0, and 0.06) with GaN. The average of the determined valence band offsets using the Zn $3d$ and Ge $3d$ bands were used to align the valence band maxima of (ZnGe)$_{0.94}$Ga$_{0.12}$N$_2$ and ZnGeN$_2$ relative to the VBM of GaN.